\documentstyle[twocolumn,prl,aps,epsfig]{revtex}
\newcommand{\beq}{\begin{eqnarray}} 
\newcommand{\eeq}{\end{eqnarray}} 
\def\lmd{\omega}
\begin{document} 
\draft 
 
\title 
{Finite Temperature Density Instability at High Landau Level Occupancy}
\author{Tudor D. Stanescu, Ivar Martin, and Philip Phillips}\vspace{.05in}

%
\address
{Loomis Laboratory of Physics\\
University of Illinois at Urbana-Champaign\\
1100 W.Green St., Urbana, IL, 61801-3080}

%

\address{\mbox{ }}
\address{\parbox{14.5cm}{\rm \mbox{ }\mbox{ }
We study here the onset of charge density wave instabilities 
in quantum Hall systems at finite temperature for Landau level
filling $\nu>4$. Specific emphasis is placed on the role of 
disorder as well as an in-plane magnetic field.
Beyond some critical value, disorder is observed to suppress 
the charge density wave melting temperature to zero.  In addition, we find that
a transition from perpendicular to parallel stripes (relative to
the in-plane magnetic field) exists when the electron gas thickness
exceeds $\approx 60$\AA. The perpendicular alignment of the stripes is
in agreement with the experimental finding that the easy conduction
direction is perpendicular to the in-plane field. }}
\address{\mbox{ }}
\address{\mbox{ }}

\maketitle
Recent experiments by Lilly and co-workers\cite{lilly} 
as well as by Du and co-workers\cite{du} have shown surprisingly that
when a 2D electron gas is subjected to perpendicular magnetic
fields of intermediate strength
 so that multiple Landau levels ($N\ge 2$) are occupied,
the resultant transport properties are strongly anisotropic.  Specifically,
the longitudinal resistivity was found to depend on the direction of the 
current at sufficiently low temperature for filling factors $\nu>4$. While the anisotropy is largest
at $\nu=9/2$ and decreases subsequently as $\nu$ is increased, the signature
anisotropy is observed to persist for filling factors up to $\nu=23/2$.
This behaviour is in contrast to what is traditionally observed
at high magnetic fields such that only the $N=0$ or $N=1$ Landau
levels are occupied\cite{fqhe,laughlin,willet}.  In addition,
subsequent experiments have shown~\cite{du2,lilly2} that the axis along which the anisotropy
is oriented can be changed by a sufficiently strong in-plane magnetic field.
Further, at $\nu=5/2$, an anisotropic state is observed only in the presence
of a parallel field.  This indicates that in the presence of an in-plane
field, the anisotropic electronic state is lower in energy than the corresponding
isotropic state at $\nu=5/2$. 
 
Anisotropic transport is inconsistent with a uniform electron
state.  As a consequence, 
charge density wave (CDW) formation has risen to the fore as the leading explanation
of these experiments\cite{lilly,du}.  In principle, the ground state 
of a 2D electron gas is expected to be 
non-uniform when the magnetic field ($B$) is sufficiently low so that no fractional
quantum Hall states are accessible but not so low that disorder destroys the 
gap established by the cyclotron frequency, $\omega_c$.  In this intermediate
range of magnetic fields, the ground state is governed by the competition between
the short-range exchange and long-range direct Coulomb interactions.  Because
these interactions are of opposite sign, the electron gas breaks into 
domains.
Several years ago,
Fogler, Koulakov, and Shklovskii\cite{kfs}(FKS) and Moessner and
Chalker\cite{mc} (MC) showed that at the Hartree-Fock level at $T=0$,
 the filling in the highest Landau level oscillates between zero and unity
with a well-defined period. Periodic stripe or bubble phases were found
to be stable as the filling in the top Landau level was varied.  
Recent numerical
simulations\cite{haldane} on small numbers of particles have also confirmed
charge density wave formation for $N\ge 2$. 
  
In this work, we focus on the role disorder and a finite in-plane
magnetic field play on the CDW instability at finite
temperature for $N\ge 2$. First, we include disorder
in an iterative screened Hartree-Fock approach that incorporates
Landau level mixing explicitly.  Second, we compare our results
with those of previous effective theories~\cite{mc,ag} for the top Landau level.
We find that the 
results from the effective theory for the width of the stripes and the melting
temperature are in good agreement with the approach used here.
We then adopt the effective interaction approach to study the role
of an in-plane magnetic field.  The key assumption is that it is the
orbital effect
of the in-plane field rather than the Zeeman energy that affects the CDW
instability.
In addition, we find that
a transition from perpendicular to parallel stripes (relative to
the in-plane magnetic field) exists when the electron gas thickness
exceeds $\approx 60$\AA. 

The starting point for our analysis is the Hartree-Fock (HF) equation,
$(H_0+\Sigma_\sigma^{\rm HF})\psi_{n\sigma x_0}({\bf r})=E_{n\sigma x_0}
\psi_{n\sigma x_0}({\bf r})$,
where $H_0$ is the non-interacting Hamiltonian for a system of electrons
in a transverse magnetic field, $\Sigma_\sigma^{\rm HF}$ is the Hartree-Fock
self-energy, $\psi_{n\sigma x_0}({\bf r})$ are the single particle
wavefunctions and $E_{n\sigma x_0}$ are the single particle energies.
This matrix equation leads naturally to Landau-level mixing when the self
energy, $\Sigma^{\rm HF}$, is non-diagonal as is the case for an inhomogeneous
system.  To solve this equation, we expand the single particle wavefunctions
\beq\label{cnn}
\psi_{n\sigma x_0}({\bf r})=\sum_{n'} C_{nn'}(\sigma,x_0)\phi_{n'x_0}
\eeq
in terms of the Landau basis,
$\phi_{nx_0}({\bf r})= \exp(-ix_0 y/\ell^2)\Phi_n(x-x_0)/\sqrt{L_y}$,
where $\Phi_n$ is a harmonic oscillator state centered at
$x_0=2\pi\ell^2 p/L_y$ with $p=0, \pm 1,\cdots$ and $L_y$ is the width of the sample
in the y-direction.  Due to computational constraints, we limit our analysis to the 
configurations with translational symmetry in one direction.  Therefore we are concerned 
only with stripe-like instabilities.
In the Landau basis, the HF equations
\beq\label{hmn}
\sum_{n'}H_{mn'}C_{nn'}(\sigma,x_0)=E_{n\sigma x_0}C_{nm}(\sigma,x_0)
\eeq
reduce to a self-consistent determination
of the expansion coefficients $C_{nn'}$ and $E_{n\sigma x_0}$.  The matrix elements
of the Hamiltonian involve a free term,
$(H_0)_{nn'}=\delta_{nn'}(\hbar\omega_c(n+1/2)-\sigma g\mu_B B/2)$,
where $\mu_B$ is the Bohr magneton and $g$ the gyromagnetic ratio
as well as contributions from direct and exchange self-energy
interactions:
\beq
\Sigma_{nn'}^H(x_0)&=&\sum_{m\sigma y_0}f(E_{m\sigma y_0})
\langle\phi_{nx_0}\psi_{m\sigma y_0}|v|
\phi_{n' x_0}\psi_{m\sigma y_0}\rangle\nonumber\\
\Sigma_{nn'}^{\rm ex}(\sigma x_0)&=&-\sum_{m y_0}f(E_{m\sigma y_0})
\langle\phi_{nx_0}\psi_{m\sigma y_0}|u|
\psi_{m\sigma y_0}\phi_{n' x_0}\rangle\nonumber
\eeq
with  $f(E_{n\sigma y_0}) = [\exp((E_{n\sigma y_0}-\mu)/T)+1]^{-1}$.
In the direct term, the Coulomb interaction, $v(q)={2\pi e^2}/{\kappa|q|}$, is
 screened only by the 
semiconductor background with static dielectric constant $\kappa$,
while the exchange term, $u(q)=2\pi e^2/\kappa\epsilon(q)|q|$,
is screened by the electron liquid with dielectric constant,
$\epsilon(q)=1-(2\pi e^2/q)\chi(q)$ \cite{mg}.

At the level
of the random-phase approximation (RPA), the exact expressions 
for this polarization effect can be calculated only for a homogeneous system. 
To extend
these results to the inhomogeneous case, we perform an average of
 $\chi$ over the period of the CDW.  This procedure should be accurate as long 
as $\chi$ is slowly varying
in the range of filling factors spanned by the period of the CDW.  
Consequently, following the result of Manolescu and
Gerhardts~\cite{mg}, we write
the inter-Landau susceptibility
as
\beq\label{chiq}
\chi(q)&=&\frac{1}{2\pi\ell^2}\sum_{n\ne n'}\left[F_{nn'}
\left(\frac{(q\ell)^2}{2}\right)\right]^2\nonumber\\
&&\times \frac1R\sum_{\sigma}\sum_{x_0=1}^R\frac{f(E_{n\sigma x_0})-f(E_{n'\sigma x_0})}
{E_{n\sigma x_0}-E_{n'\sigma x_0}}
\eeq
where $R$ is the number of guiding centers within a period,
\beq\label{Lag}
F_{nn'}(z)=\left(\frac{n'!}{n!}\right)^\frac12 z^{(n-n')/2} e^{-z/2} L_{n'}^{n-n'}(z)
\eeq
and $L_n^m(z)$ is an associated Laguerre polynomial. 

In the presence of disorder, the energy levels acquire finite width.  
Experimentally, the broadening parameter, $\Gamma$, has been found to scale as 
$\Gamma = \Gamma_0\sqrt{B}$.  Phenomenologically, this behavior can be incorporated 
into our scheme by
using the spectral function in the form~\cite{mol}
\beq
\rho_\sigma(n, y_0,\omega) = \frac1{\sqrt{\pi}}\frac\hbar\Gamma
\exp\left[-\frac{(\hbar\omega - E_{n \sigma y_0})^2}{\Gamma^2}\right].
\eeq
As a consequence the Fermi functions $f(E_{n\sigma y_0})$ in the above
equations need to be replaced by the filling factors 
\beq
\nu_{n\sigma y_0} = \int_{-\infty}^{\infty}d \omega f(\hbar \omega) \rho_\sigma(n,y_0; \omega).
\eeq

We solve Equations~(\ref{hmn}--\ref{chiq}) in real space starting from 
a non-interacting initial guess for the energy levels. To evaluate 
the inhomogeneous screening,
we chose a grid composed of 16 to 32 points per period of the CDW. 
We included as many as 10 Landau levels.  The parameters used 
in the computation (which are specialized to GaAs) are as
 follows: $m^\ast=0.067m_e$, $g=0.4$, $\kappa=12.7$,
and $n=2.4\times 10^{11} cm^{-2}$.  With these parameters,
$\hbar \omega_c = 20 B$, for the cyclotron gap, 
$e^2/\kappa \ell = 51 \sqrt{B}$, for the strength of Coulomb interaction, 
and $0.133B$ for
bare spin splitting.  
Here, the energies are measured in Kelvin and magnetic field, B, in Tesla.  
All of the computations were performed for a fixed number of electrons,
with the chemical potential, $\mu$, being determined self-consistently.

In order to obtain the CDW versus homogeneous phase diagram, we must first determine
the optimal period of the CDW.  The optimal 
period yields the highest melting temperature for a given average filling.
We find that this wavelength coincides with the period of the CDW with the lowest 
free energy.  The results for the optimal period, $\lambda_1$, 
at half filling in terms of magnetic length $\ell$
are shown below:

\vspace{2mm}
\begin{tabular}{|c||c|c|c|c|} \hline
$\ \ \ \ \nu\ \ \ \ $   &$\ \ \ 4.5\ \ \ $ 	&$\ \ \ 5.5\ \ \ $	&$\ \ \ 6.5\ \ \ $	&$\ \ \ 7.5\ \ \ $\\ \hline 
$\lambda_1$		&5.84	&5.90	&6.93	&6.99\\ \hline 
$\lambda_2$             &6.13	&6.19	&7.18	&7.22\\ \hline 
$\lambda_3$	&6.04	&6.04	&7.14	&7.14\\ \hline   
\end{tabular}
\vspace{2mm}

The error in the determined values is on the order of $1\%$.  From the 
theory of FKS, it is expected that the period at half-filling, $\lambda_{3}$,
of the CDW should only depend on $N$.  Indeed, we observe minor variation for different spin polarizations,
and the periods we obtain are quite close to 
the FKS value, $\lambda_{3} = 2.7\sqrt{2N+1}$.
Within the same Landau level, we find however, a reduction in $\lambda$ away from half-filling.  
For example, at filling fractions 0.2 and 0.8, the reduction is on 
the order of $4-5\%$.  The values of $\lambda_2$ were obtained
from an effective theory for the top Landau level~\cite{mc}.  These values are also
in close agreement with the theoretical estimate and the results from the
 numerical scheme outlined above.

\begin{figure}
\begin{center}
\epsfig{file=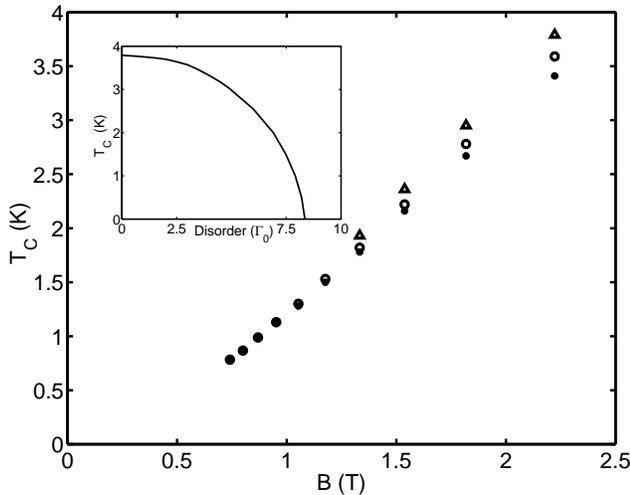, height=6.5cm}
\caption{ Melting temperatures for CDW formation
at half-filling as a function of the perpendicular
magnetic field for three cases: 1) $\triangle$ represent the melting temperatures
from the screened Hartree-Fock approach, 2) $\circ$ were calculated from 
Eq.~(\protect\ref{tc}) for zero thickness of the 2D layer, and 3) $*$ the same method
but with an rms thickness of $100$\AA.
The inset shows the melting temperature as a function of disorder for $\nu=4.5$.}
\label{fig1}
\end{center}
\end{figure}

Shown in Fig.~(\ref{fig1}) are the melting temperatures at half
filling as a function of magnetic field.  The triangles represent the
results for the screened Hartree-Fock approach.  
As mean-field estimates these values
 will certainly be reduced by quantum fluctuations~\cite{fradkin}
and hence set an upper bound for the temperature at which 
a  charge instability occurs.  These values do, however, conform to the experimental
trend that the temperature at which the anisotropy is observed decreases
as $\nu$ increases.   Also shown in Fig.~(\ref{fig1}) are the melting
temperatures computed using the effective theory~\cite{mc} for the top
Landau level.  The circles represent melting temperatures at half-filling
and the asterisks to the corresponding quantity for an electron gas
with root-mean-square (rms) thickness
of $100$\AA.  As is evident, all three estimates for the CDW 
melting temperatures are in close agreement suggesting that the approximations
incurred in the effective theories are consistent.  
The inset in Fig.~(\ref{fig1}) shows how
the disorder as measured by $\Gamma_0$ attenuates CDW formation at
$\nu = 4.5$.  As is evident, a
 critical value of the disorder,  
$\Gamma^c_0 \approx 8.3$, exists above which the CDW phases desist. 
Experimentally, the disorder parameter $\Gamma_0$ can be estimated, for example, from
the lowest magnetic field for which de Haas-Van Alphen oscillations are
observed.  In very clean 
samples\cite{lilly}, the oscillations disappear below 50 mT.  Assuming that the 
oscillations disappear when $2\Gamma\sim\hbar\omega_c$, we estimate
that  $\Gamma_0 \sim 2.2 [K/T^{1/2}]$.  We find, that for the range of fillings studied ($\nu\leq15/2$) the critical value $\Gamma_0^c>2.2$.

To understand the role of an in-plane field, we adopt the simpler approach 
in which the interactions in the top level
are treated within an effective theory\cite{mc,kfs,ag}.  
This simplification is warranted as we have seen that the results of the effective
theories are in agreement with the more elaborate scheme described above.
In the spirit of the work of 
Fukuyama, Platzman and Anderson~\cite{fpa}, we work within
an effective Ginsburg-Landau theory and write the 
melting temperature for a CDW in a 2D electron
gas in a perpendicular magnetic field as
\beq\label{tc}
T_c=\bar{\nu}(1-\bar{\nu})|U(\vec Q_0)|,
\eeq
with $\bar{\nu}$ is the fractional occupancy in the top Landau level
and $U(\vec Q_0)$ is the effective Hartree-Fock potential evaluated
at the momentum $\vec Q_0$ where the potential is minimized.  

The primary hurdle in computing $T_c$ is the
effective Hartree-Fock potential $U({\bf Q}_0)$.  As the spin degrees
of freedom do not couple to the stripe orientation, the spin-splitting
induced by the in-plane field cannot account for the experimental
observations.  We focus then on the orbital
effect of the in-plane field.  To this end, we consider our quantum
Hall system to have a finite thickness in the direction normal
to the 2D plane.  Due to this finite 
thickness, the one particle wave functions couple to the in-plane component 
of the magnetic field $B_{||}$. In what follows, we will neglect the 
contribution due to the coupling of $B_{||}$ to the electron spin. 
Let us consider a system of electrons moving in the $x-y$ plane in
the presence of a tilted magnetic field ${\bf B}=(B \tan\theta, 0, B)$. 
We assume that the confining 
potential in the $z$ direction is harmonic $V(z) = m^\ast 
\Omega^2z^2/2$, with
a characteristic 
frequency determined from the rms thickness in the z-direction,
$\Omega=\hbar/2mL_z^2$. While the actual confining potential
differs from the parabolic form chosen here, the main features
of the finite thickness should be correctly captured by 
appropriately choosing the rms width.  By solving the Schr\"odinger equation in the gauge 
${\bf A} = (0,-zB\tan\theta+xB,0)$, we obtain the one-particle wave functions for this system:
\beq\label{wf}
\phi_{nx_0}({\bf r}) &=& \frac{1}{\sqrt{L_y}} e^{-\frac{ix_0 y}{\ell^2}}
\Phi_0^{\lmd_{+}}((x-x_0)\sin\tilde{\theta}  + z\cos\tilde{\theta} ) \nonumber \\
&\times&\Phi_n^{\lmd_{-}}( (x-x_0)\cos\tilde{\theta} -z \sin\tilde{\theta} ) 
\eeq
where $\Phi_n^{\omega}$ is the harmonic oscillator wave 
function corresponding to the frequency $\omega$ and $\tilde{\theta}$ is given by
$\tan\tilde{\theta} =\tan \theta\omega_c^2/(\lmd^2_{+} -\omega^2_c)$.
The frequencies $\lmd_{\pm}$ 
\beq\label{lamb}  
\lmd^2_{\pm} = \frac12 (\Omega^2 + \frac{\omega_c^2}{\cos^2\theta}) \pm 
\sqrt{\frac1{4}(\Omega^2 - \frac{\omega_c^2}{\cos^2\theta})^2 + 
 \Omega^2\omega_c^2 \tan^2 \theta}\nonumber
\eeq
depend on the strength and
orientation of the magnetic field and on the confining potential.
In the absence of $B_{||}$, $\lmd_+$ reduces to 
$\Omega$ and $\omega_-$ to the cyclotron frequency $\omega_c$.

Using the one-particle states Eq.~(\ref{wf}), we are able 
now to compute the matrix elements $\langle \phi_{nx_0}
|e^{i{\bf Q}\cdot{\bf r}}| \phi_{my_0} \rangle$ needed for 
the determination of the effective potential $U({\bf Q}_0)$.
Explicitly, we write the effective potential as a sum of a Hartree term 
\beq\label{uh}
U_H(q_x,q_y) = \frac{e^2}{\kappa\epsilon(q_x,q_y) l_0^2}
 \int_{-\infty}^{\infty} \frac{dq_z}{\pi} \frac 1{q_{||}^2 
+ q_z^2} \left[F_{nn}^{\theta}({\bf q})\right]^2\nonumber
\eeq
and a Fock term 
\beq\label{uf}
U_F(q_x,q_y) = - 2\pi l_0^2 \int \frac{dk_x dk_y}{(2\pi)^2} U_H(k_x,k_y)
             e^{i(k_x q_y - k_y q_x)l_0^2}\nonumber
\eeq
where $ F_{nn'}^{\theta}({\bf q})$ is the analog of Eq.~(\ref{Lag}) 
for the case of a tilted field and is given by
\beq\label{Fnn}
F_{nn'}^{\theta}(q_x,q_y,q_z) = \left(\frac{n'!}{n!}\right)^{\frac1 2}
\left(\frac{\alpha^2}{2}\right)^{\frac{n-n'}{2}} e^{-\frac{\gamma^2}{4} -
\frac{\alpha^2}{4}} L_{n'}^{n-n'}(\frac{\alpha^2}{2})\nonumber
\eeq
with
\beq
\gamma^2 &=& (q_z \cos\tilde{\theta} + q_x \sin\tilde{\theta})^2 l_+^2 
+ q_y^2 \sin^2\tilde{\theta} l_0^4/l_+^2 \nonumber \\
\alpha^2 &=& (q_x \cos\tilde{\theta} - q_z \sin\tilde{\theta})^2 l_-^2
+ q_y^2 \cos^2\tilde{\theta} l_0^4/l_-^2.
\eeq
The dielectric constant $\epsilon(q_x,q_y)$ is calculated 
in the RPA approximation (see Eq.~(\ref{chiq})). 
In this case, we used the non-interacting 
energies $E_{n\sigma}^0 = \hbar\omega_c(n + \frac12) - 
\frac{\sigma}{2} g \mu_B B$ instead of the self-consistent 
energies $E_{n\sigma x_0}$ and the {\bf q} dependence enters
through the angle dependent function $F_{nn'}^{\theta}(q_x,q_y,0)$. 

\begin{figure}
\begin{center}
\epsfig{file=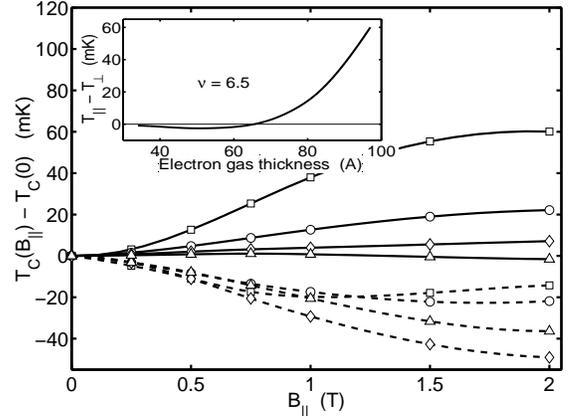, height=5.5cm}
\caption{ Change in melting temperature of the CDW stripe phases as a result
of a parallel magnetic field for an rms thickness of 100\AA.  The solid curves correspond to stripes
oriented parallel to the field and the dashed lines to a perpendicular alignment
with respect to the field.  Four filling factors are shown: $\Box$, o,
$\Diamond$, and $\triangle$ correspond to $\nu=$4.5, 5.5, 6.5, and 7.5,
respectively. The inset shows the difference in the melting temperature 
between parallel and perpendicular stripes as
a function of the thickness of the electron gas in \AA.}
\label{fig2}
\end{center}
\end{figure}

Using this method, we minimized the effective potential for two orientations
of the stripes: perpendicular and parallel to the in-plane magnetic field.
In general, the minimum value of the effective potential is displaced
both vertically in energy and horizontally in wavevector. With extreme
consistency, we observed that the wavevector for stripes
oriented parallel to the in-plane field always exceeded the wavevector for
those perpendicular to the stripes.  This difference was not more
than 4$\%$.  Our results for the change in the melting
temperature ($T_c$) 
 as  a function of an in-plane field for an rms thickness of 100\AA are
shown in Fig.~(\ref{fig2}).  
The solid curves correspond to a parallel 
orientation of the stripes (with respect to the in-plane field)
while the dashed curves denote the perpendicular 
orientation of the stripes.  As is evident,
the melting temperature for stripes oriented
parallel to the field is consistently higher than 
$T_c$ for the perpendicular
orientation as $\nu$ is incremented.  However, as shown in the inset, 
this trend reverses as the rms thickness
is reduced below 65.3\AA for $\nu=6.5$.  Below this thickness, the
difference in melting temperature between
perpendicular and parallel stripes is $O(1mK)$
for a magnetic field tilted at $45^o$. This temperature
difference directly translates into an energy difference per electron
between the two phases at low temperatures\cite{kfs}. To determine which
phase is preferred, we must compare the total energy difference
with the temperature.  As the temperature is lowered below $T_c$,
the correlation length of the ordered charge density wave domains
grows until perfect orientational order is reached at the
Kosterlitz-Thouless temperature. For sufficiently
large domains, even miniscule energy differences per electron
can lead to orientational order and hence anisotropic transport.
Hence, we predict that for thin electron gases ($\alt$60\AA),
the perpendicular orientation is preferred, whereas for thicker electron
gases ($\agt$60\AA) parallel stripes dominate.  Interestingly
enough, Eisenstein\cite{lilly3} places the electron gas thickness at
58\AA\, and experiments\cite{lilly2} identify the direction of the in-plane field as being
the high resistive direction.  This experimental result is consistent
with our finding that perpendicular stripes are energetically favoured
for thin electron gases. 

\acknowledgements
We thank E. Fradkin, C. Chamon, S. Girvin, T. Jungwirth, A. MacDonald, and A.
Manolescu for
helpful criticism, M. Fogler and J. Eisenstein for alerting us to the role of 
strong confinement of the electron gas, and A. Koulakov for
a discussion on $T_{KT}$ in the stripe phase and the ACS Petroleum
Research Fund and the NSF grant No. DMR98-96134. Corroborating results can 
also be found in cond-mat/9905353.

\end{document}